\documentclass[final,5p,times,twocolumn]{elsarticle}
\usepackage{amsmath,amssymb}
\usepackage{graphicx}
\usepackage{units}
\usepackage{hyperref}

\journal{Nuclear Physics B}

\newcommand{\e}{\mathrm{e}}

\begin{document}

\begin{frontmatter}

  \title{Standard Model Higgs boson mass from inflation}

  \author[MPI,INR]{Fedor L. Bezrukov}
  \ead{Fedor.Bezrukov@mpi-hd.mpg.de}
  \author[EPFL]{Amaury Magnin}
  \ead{Amaury.Magnin@epfl.ch}
  \author[EPFL]{Mikhail Shaposhnikov}
  \ead{Mikhail.Shaposhnikov@epfl.ch}

  \address[EPFL]{
    Institut de Th\'eorie des Ph\'enom\`enes Physiques,
    \'Ecole Polytechnique F\'ed\'erale de Lausanne,
    CH-1015 Lausanne, Switzerland}
  \address[MPI]{
    Max-Planck-Institut f\"ur Kernphysik,
    PO Box 103980, 69029 Heidelberg, Germany}
  \address[INR]{
    Institute for Nuclear Research of Russian Academy of Sciences,
    Prospect 60-letiya Oktyabrya 7a,
    Moscow 117312, Russia}

  \date{March 3, 2009}

  \begin{abstract}
   We analyse one-loop radiative corrections to the inflationary
potential in the theory, where inflation is driven by the Standard
Model Higgs field. We show that inflation is possible provided the
Higgs mass $m_H$ lies in the interval $m_{\rm min}< m_H < m_{\rm
max}$, where 
$m_{\rm min} = [136.7 + (m_t - 171.2)\times 1.95]\,\unit{GeV},~~
 m_{\rm max} = [184.5 + (m_t - 171.2)\times 0.5]\,\unit{GeV}$ 
 and $m_t$ is the mass of the top quark. In the renormalization scheme
associated with the Einstein frame the predictions of the spectral
index of scalar fluctuations and of the tensor-to-scalar ratio
practically do not depend on the Higgs mass within the admitted region
and are equal to $n_s=0.97$ and $r=0.0034$ correspondingly.
      \end{abstract}


\end{frontmatter}
\section{Introduction}
During the last decades a growing number of connections between
cosmology and particle physics were established.  However, finding a
relation of cosmological inflation to low energy particle theory is a
difficult task. In many models inflation is driven by some new physics
at large energies which is not connected to the scale of the Standard
Model (SM).  In \cite{Bezrukov:2007ep} it was  suggested that the SM
Higgs boson can play the role of the inflaton. At first sight, the
properties of the electroweak Higgs boson (with the quartic coupling
$\lambda\sim 0.1$ and the mass $m_H\sim\unit[100]{GeV}$) are very far
from those required for the inflaton field \cite{Linde:1983gd}  (in
the simplest $m^2\phi^2+\lambda\phi^4$ model  the typical choice of
parameters, leading to successful inflation, is $\lambda \sim
10^{-13}$, $m \sim \unit[10^{13}]{GeV}$). Nevertheless, addition of
the non-minimal coupling of the Higgs fields to the Ricci scalar
changes the situation.  As was shown in \cite{Bezrukov:2007ep}, the SM
action with gravity included
\begin{equation}
  \label{SMJordan}
  S_J = S_\mathrm{SM} +
    \int d^4x\sqrt{-g}\, \left(
      - \frac{M^2}{2}R - \xi\Phi^\dagger\Phi R
    \right)
\end{equation}
naturally leads  to inflation.  Here $S_\mathrm{SM}$ stands for the SM
action, $M$ is  a mass parameter, nearly equal to the Planck mass in
our case, $R$ is the scalar curvature, $\Phi$ is the Higgs doublet,
and $\xi$ is the new coupling constant.

The fact that non-minimal coupling of the scalar field relaxes the
requirement for the smallness of the quartic coupling, and also
suppresses the generation of the tensor modes during inflation, was
known for quite a long time
\cite{Spokoiny:1984bd,Salopek:1988qh,Fakir1990,Komatsu:1999mt,%
Tsujikawa:2004my,Barvinsky1994,Barvinsky1998}.  Basically, if
non-minimal coupling is present, the parameter that fixes the
normalization of the CMB fluctuations is not the scalar self-coupling,
but the combination $\lambda/\xi^2$. It is this point which allows the
SM Higgs boson to be the inflaton at the same time.

The study of \cite{Bezrukov:2007ep} was based on the classical scalar
potential in the theory (\ref{SMJordan}). It was argued there that the
radiative corrections do not spoil the flatness of the potential,
necessary for inflation. In
Refs.~\cite{Bezrukov:2008cq,Shaposhnikov:2008rc} it was conjectured
that all the results of the tree analysis remain true if the Higgs
mass lies in the interval\footnote{These specific numbers should be
taken with a grain of salt, as they were quoted in
\cite{Shaposhnikov:2007nj} on the basis of compilation of previous
computations and do not take into account the progress made in
experimental determination of electroweak parameters.} $m_H
\in[129,189]\,\unit{GeV}$, corresponding to the situation when the
Standard Model remains a consistent quantum field theory up to the
inflation scale $M_P/\xi$, or, to be on a safer side, all the way up
to the Planck scale $M_P$.

The aim of this Letter is the analysis of the  renormalization group
improved effective potential for Higgs-inflaton. We will show that
inflation is possible in the SM if the Higgs mass lies in the interval
$m_{\rm min}< m_H < m_{\rm max}$, somewhat exceeding the range in
which the SM can be valid up to the Planck scale, in accordance with
our previous expectations.

The Letter is organized as follows. In Sec. \ref{sec:tree} we review
the Higgs-inflaton scenario and introduce the notations. In Sec.
\ref{sec:rg} we construct the renormalization group improved effective
potential and discuss possible renormalization prescriptions for its
computation. We also identify there an error made in a previous
attempt \cite{Barvinsky:2008ia} to include radiative corrections to
Higgs-inflation.  In Sect. \ref{sec:num} we present the numerical
results. Sect. \ref{sec:conc} is conclusions.

\section{Inflation in tree approximation}
\label{sec:tree}
The simplest way to analyse the inflation in the model
(\ref{SMJordan}) is to make the conformal transformation to the
``Einstein frame'', where the gravitational term takes its usual
form.  This is achieved by rescaling the metric by the conformal
factor $\Omega$
\begin{equation}
  \label{Omega}
  g_{\mu\nu} \to \hat{g}_{\mu\nu} = \Omega^2 g_{\mu\nu}
  \;,\quad
  \Omega^2 = \frac{M^2+\xi h^2}{M_P^2}
  \;,
\end{equation}
where $M_P\equiv1/\sqrt{8\pi G_N}=\unit[2.44\times10^{18}]{GeV}$ is
the reduced Planck mass, and $h$ is the unitary gauge Higgs
$\Phi(x)=\frac{1}{\sqrt{2}}{0\choose v+h(x)}$.  Then, with
redefinition of the field $h\to\chi$
\begin{equation}
  \label{dchidh}
  \frac{d\chi}{dh} = \sqrt{\frac{\Omega^2+6\xi^2h^2/M_P^2}{\Omega^4}}
  \;,
\end{equation}
we get the action with usual gravity and canonically normalised scalar
field $\chi$ with potential
\begin{equation}
  \label{Uexact}
  U(\chi) =
    \frac{1}{\Omega^4\left[ h(\chi) \right]} 
    \frac{\lambda}{4}\left[h^2(\chi) -v^2\right]^2
  \;.
\end{equation}
For large $\xi$ the approximate solution for (\ref{dchidh}) is 
\begin{equation}
  \label{chi(h)}
  \chi \simeq \left\{
    \begin{array}{l@{\qquad\text{for}\quad}l}
      h                                    & h<\frac{M_P}{\xi}
      \;,\\
      \sqrt{\frac{3}{2}}M_P\log\Omega^2(h) & \frac{M_P}{\xi} < h
      \;.
    \end{array}
  \right.
\end{equation}
Therefore, the potential coincides with the standard one for small
fields.  At the same time, for large fields it becomes exponentially
flat
\begin{equation}
  \label{U(chi)}
  U(\chi) \simeq \frac{\lambda M_P^4}{4\xi^2}
    \left(
      1-\e^{-\frac{2\chi}{\sqrt{6}M_P}}
    \right)^{2}
  \;.
\end{equation}
The inflation in the Einstein frame\footnote{The same results can be
obtained in the Jordan frame
\cite{Tsujikawa:2004my,Makino1991,Fakir:1992cg}.} can be analysed by
the usual means \cite{Bezrukov:2007ep,Bezrukov:2008ut,Linde:2007fr}.  
One has a slow roll inflation ending at
$h_\mathrm{end}\simeq(4/3)^{1/4}M_P/\sqrt{\xi}$, with the WMAP scale
perturbations exiting the horizon $N\simeq59$ e-foldings earlier at
$h_\mathrm{WMAP}\simeq9.14M_P/\sqrt{\xi}$.  The normalization of the
CMB perturbations leads to the requirement
\begin{equation}
  \label{WMAPnorm}
  \xi \simeq \sqrt{\frac{\lambda}{3}}\frac{N_\mathrm{WMAP}}{0.0276^2}
      \simeq  44700\sqrt{\lambda}
  \;.
\end{equation}
The spectral index and the tensor-to-scalar ratio are  $n_s\simeq
0.97$ and $r \simeq 0.0034$, which lies well within the WMAP5 limits
\cite{Komatsu:2008hk}.

\section{Renormalization group and effective potential}
\label{sec:rg}
Following \cite{Bezrukov:2007ep}, we adopt the following procedure for
computation of quantum corrections to the effective potential. First,
we rewrite the theory in the Einstein frame with the use of the
equations given in the previous Section. Second, we determine the
particle masses (W, Z, Higgs, and t-quark) as a function of the
background field $\chi$:
\begin{gather}
  \label{masses}
  m_W^2 = \frac{g^2 h^2}{4\Omega^2}
  \;,\;
  m_Z^2 = \frac{(g^2+g'^2) h^2}{4\Omega^2}
  \;,\;\\
  m_H^2 = \frac{d^2U}{d\chi^2}
  \;,\;
  m_t^2 = \frac{y_t^2 h^2}{2\Omega^2}
  \;.
  \nonumber
\end{gather}
Here $g$, $g'$, $g$ are the electroweak SU(2), U(1) and strong SU(3)
coupling constants, $y_t$ is the top quark Yukawa coupling. Note that
the only difference from the flat space case is the presence of the
conformal factor $\Omega$ in the denominators of the masses. This is
just the result of the transformation to the Einstein frame. Finally,
we compute the radiative corrections with the use of the standard
formula of \cite{Coleman:1973jx} (cf.\ \cite{Ford1993,Ford1992}):
\begin{equation}
  \label{dU1loop}
  \delta U = \frac{6m_W^4}{64\pi^2}\log\frac{m_W^2}{\mu^2}
    + \frac{3m_Z^4}{64\pi^2}\log\frac{m_Z^2}{\mu^2} 
    - \frac{3m_t^4}{16\pi^2}\log\frac{m_t^2}{\mu^2}
  \;,
\end{equation}
where $\mu$ is the normalization scale\footnote{In the ${\overline{\rm
MS}}$ subtraction scheme $\log(x)$ should be replaced by $\log(x)
-3/2$ for t-quark and $\log(x)-5/6$ for gauge bosons.}. Note that we
omitted the contribution from the Higgs field itself, since it is
exponentially suppressed at large field values and thus can be safely
neglected for analysis of inflation.

This procedure may be contrasted with that of \cite{Barvinsky:2008ia}.
These authors suggested to use the Jordan rather than Einstein frame
for the computation of radiative corrections and made a transition to
the Einstein frame of the 1-loop effective potential. This leads to
the same structure of radiative corrections as in Eq.~(\ref{dU1loop})
but with replacement $\mu\to\mu/\Omega$. 

Let us elaborate more on the difference between two prescriptions. For
this end we replace the normalization point $\mu$ by an effective
ultraviolet ``cut off''.  Then two choices are possible with the cut
off proportional to:
\begin{center}
  \begin{tabular}{l|c|c}
                   & I \cite{Bezrukov:2007ep}& II \cite{Barvinsky:2008ia}
    \\\hline
    Jordan frame   & $M_P^2+\xi h^2$          & $M_P^2$
    \\\hline
    Einstein frame & $M_P^2$                  & \(\displaystyle
                                      \frac{M_P^4}{M_P^2+\xi h^2}
                                                    \)
  \end{tabular}
\end{center}
Both choices correspond to imposing a \emph{field-dependent} cut off
in one or another frame. It is hard to say, which prescription should
be used without knowledge of the behaviour of the quantum theory at
the Planck scales. In fact, the prescription of \cite{Bezrukov:2007ep}
can be justified by the ideas of exact quantum scale invariance,
discussed in
\cite{Shaposhnikov:2008xb,Shaposhnikov:2008xi,Shaposhnikov2008}.  In
these papers it was proposed that all dimensional parameters,
including the Planck mass, are generated by spontaneous breaking of
exact scale invariance by an additional dilaton field. The SI-GR
prescription of \cite{Shaposhnikov:2008xi} exactly corresponds to the
suggestion which was made in \cite{Bezrukov:2007ep}. Though the
choices of \cite{Bezrukov:2007ep} and \cite{Barvinsky:2008ia} are not
physically equivalent, we will show that after the renormalization
group improvement the predictions for the Higgs mass are \emph{nearly
the same.}

The one loop modification of the potential (\ref{dU1loop}) is a good
approximation, if the logarithms remain small. However, if one uses
the physical values of coupling constants, Higgs, $W,Z$ and top
masses, which are defined at the electroweak scale, the logarithms are
large in the inflationary region. Therefore, to connect the potential
at inflation with the low energy  parameters, one should apply the
renormalization group procedure. This was not done in
\cite{Barvinsky:2008ia}, which resulted in erroneous conclusions.

The one-loop renormalization group equations in the curved space are
(no graviton loops are included)
\cite{Yoon1997,Ford1992,Ford1993,Buchbinder1992}:
\begin{align}
  \label{betag}
  16\pi^2\frac{dg}{dt}       &= -\frac{19}{6}g^3 \;,\\
  16\pi^2\frac{dg'}{dt}      &= \frac{41}{6}g'^3 \;,\\
  16\pi^2\frac{dg_3}{dt}     &= -7g_3^3 \;,\\
  16\pi^2\frac{dy_t}{dt}     &= \frac{9}{2}y_t^3-8g_3^2y_t
                                -\frac{9}{4}g^2y_t-\frac{17}{12}g'^2y_t \;,
\\
  \label{betalambda}
  16\pi^2\frac{d\lambda}{dt} &= 24\lambda^2+12\lambda y_t^2
                                -9\lambda(g^2+\frac{1}{3}g'^2) \notag\\
                             &\hphantom{=}
                                -6y_t^4 
                                +\frac{9}{8}g^4+\frac{3}{8}g'^4
                                +\frac{3}{4}g^2g'^2 \;,\\
  \label{betaxi}
  16\pi^2\frac{d\xi}{dt}     &= \left(\xi+\frac{1}{6}\right)\left(
                                  12\lambda+6y_t^2
                                  -\frac{9}{2}g^2-\frac{3}{2}g'^2
                                \right) \;,
\end{align}
where $t\equiv\log\mu/M_Z$. 

The solution of these equations can be plugged in the expression for
the effective potential (as usual, the $\mu$-dependent constants
should be substituted only in the tree level part)
\begin{multline}
  U_\mathrm{eff}(\chi,\mu) = U+\delta U \\
  = \frac{\lambda(\mu)}{4\xi^2(\mu)} f(\chi)
    + s(g,g',g_3,y_t) f(\chi) \log\left(\frac{m_t^2}{\mu^2}\right)
  \\
    + \text{$\mu$-independent terms}
  \;.
\end{multline}
Note that the function
\begin{equation}
  f(\chi)= M_P^4
  \left(
    1-\e^{-\frac{2\chi}{\sqrt{6}M_P}}
  \right)^{2}
\end{equation}
is in fact the same in the tree level term and one loop contributions
(compare (\ref{U(chi)}) with (\ref{dU1loop}), (\ref{masses})), and
function $s(g,g',g_3,y_t)$ can be read of (\ref{dU1loop}),
(\ref{masses}).

The dependence of the effective potential on $\mu$ is artificial. To
be more precise, $U_\mathrm{eff}(\chi,\mu)$ does not depend on $\mu$
at its extrema (in other points contributions the field
renormalization must be taken into account). In our case, the
potential becomes constant at $\chi\to\infty$, and, therefore, it
should not depend on $\mu$ in this region (in other words, the energy
density during inflation is a physical quantity and thus is
$\mu$-independent). With the use of
Eqns.~(\ref{betalambda},\ref{betaxi}) one can easily check that this
is indeed the case for both prescriptions discussed above,
\begin{equation}
  \frac{d}{d\mu}\left[
    \frac{\lambda(\mu) M_P^4}{4\xi^2(\mu)}+ \delta U
  \right] =0
  \;.
\end{equation}
The running of $\xi$ is essential for this result.

As far as the potential is $\mu$-independent, we can choose the most
convenient value of $\mu$.  The obvious choice is to take $\mu$ to
make the logarithms vanish \cite{Stevenson:1981vj}
\begin{equation}
  \label{renormcondition}
  \mu^2 = \kappa^2 m_t^2(\chi) 
    = \kappa^2\frac{y_t(\mu)^2}{2}\frac{M_P^2}{\xi(\mu)}
      \left(
        1-\e^{-\frac{2\chi}{\sqrt{6}M_P}}
      \right)
  \;.
\end{equation}
Here $\kappa$ is some constant of order one, introduced to imitate
difference between $m_t$, $m_W$, $m_Z$, and also account for
$\mu$-independent terms that were dropped in (\ref{dU1loop}).  Then
the final improved potential is given by the formula (\ref{U(chi)}),
where $\lambda$ and $\xi$ are taken at the scale $\mu$, determined by
(\ref{renormcondition}). The parameter $\mu$ varies in a finite
interval, $0< \mu < \mu_{\rm max}$, corresponding to the $\chi$ change
from $0$ to $\infty$.

Making the analysis for the prescription of \cite{Barvinsky:2008ia} is
also simple, and boils down to just taking another value for $\mu$:
\begin{equation}
  \label{renormconditionBarv}
  \mu^2 = m_t^2(\chi)\Omega(\chi)^2 
    = \frac{y_t(\mu)^2}{2}\frac{M_P^2}{\xi(\mu)}
      \left(
        \e^{\frac{2\chi}{\sqrt{6}M_P}}-1
      \right)
  \;.
\end{equation}
Once the potential is determined, one can carry out the usual analysis
of the slow-roll inflation, fixing $\xi$ from COBE or WMAP
normalization, and calculating spectral index $n_s$ and tensor to
scalar ration $r$, in complete analogy with
\cite{Bezrukov:2007ep,Bezrukov:2008ut}. The only technical point here
is that it is easier to use $\mu$ as an independent variable instead
of $\chi$. The advantage is that no inversion of
Eqns.~(\ref{renormcondition}, \ref{renormconditionBarv}) is required.

\section{Numerical results}
\label{sec:num}
We solve the equations (\ref{betag})-(\ref{betaxi}) with the initial
conditions
\begin{gather*}
  \frac{g^2}{4\pi}=0.034
  \;,\;
  \frac{g'^2}{4\pi}=0.010
  \;,\;
  \frac{g_3^2}{4\pi}=0.13
  \;,\;\\
  y_t\frac{v}{\sqrt{2}}=\unit[171.2]{GeV}
  \;,\;
  \sqrt{2\lambda} v=m_H
  \;,\;
  \xi=\xi_0
\end{gather*}
at $\mu=M_Z$.  Here $v=\unit[246.22]{GeV}$ and the central value of
the mass of t-quark is specified for concreteness.  With this solution
we obtain the RG improved potential, which is  then used for
computation of the parameters of inflation. We take $\kappa=1$.

We find that inflation can take place provided the Higgs mass lies in the
interval 
\begin{align}
  \nonumber
  &m_{\rm min}< m_H < m_{\rm max}\;,\\ 
  \label{mhiggs}
  m_{\rm min} =& [136.7 + (m_t - 171.2)\times1.95]\,\unit{GeV}\;,\\ 
  m_{\rm max} =& [184.5 + (m_t - 171.2)\times0.5]\,\unit{GeV}\;.
  \nonumber
\end{align} 
If the mass is smaller than $m_{\rm min}$, the slope of the effective
potential for large field values becomes negative, making inflation
impossible. If the mass is larger than $m_{\rm max}$, the value of
$\mu_{\rm max}$, corresponding to inflationary stage is close to the
Landau pole in $\lambda(\mu)$, making the theory strongly coupled. The
specific numbers in (\ref{mhiggs}) correspond to $\mu_\mathrm{max}$
coinciding with the Landau pole for $\lambda$.  More elaborate
definitions of the applicability of the perturbative theory may be
introduced (like $\lambda(\mu_{\rm max})\sim1$), and lead to slightly
smaller $m_\mathrm{max}$.

\begin{figure}
  \centering
  \includegraphics[width=\columnwidth]{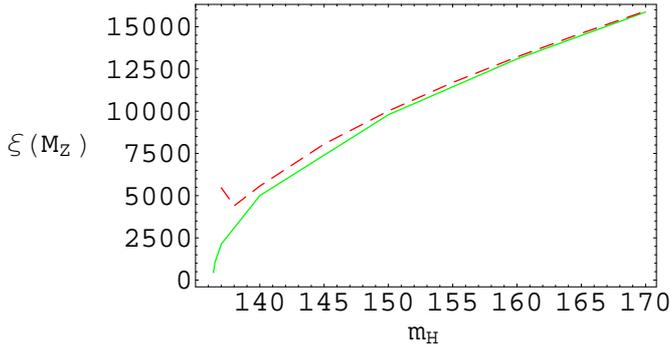}  
  \caption{Non-minimal coupling parameter $\xi$ as a function of the Higgs
    mass $m_H$.  Solid line is for the choice I, dashed---for the
    choice II.}
  \label{fig:xi}
\end{figure}

The value of $\xi$ (at the $M_Z$ scale), leading to proper CMB
normalization, is presented in Fig.~\ref{fig:xi}.  As expected,
smaller $\xi$ correspond to smaller Higgs masses,
cf.~(\ref{WMAPnorm}).  The small rise in $\xi$ at small $m_H$ for the
prescription of \cite{Barvinsky:2008ia} corresponds to the potential
which starts to decrease at high field values.
Note, that for the $\lambda$-$\xi$ relation the approximate formula
(\ref{WMAPnorm}) can still be used, (except for the Higgs masses very
close to the boundaries of the allowed region).  Of course, $\xi$ and
$\lambda$ in (\ref{WMAPnorm}) should be calculated then at the scale
$\mu$, corresponding to inflation.

\begin{figure}
  \centering
  \includegraphics[width=\columnwidth]{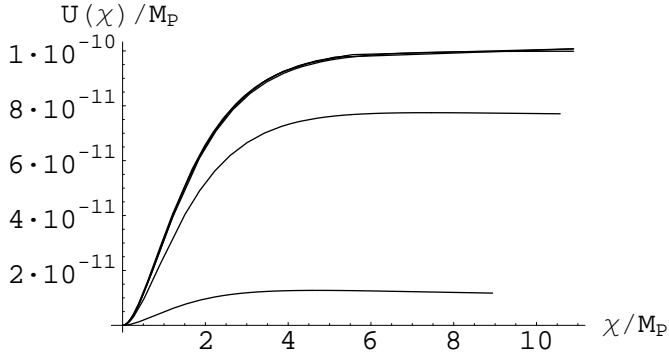}
  \caption{RG improved effective potential.  The four nearly coinciding
    upper lines correspond to the choice I and
    $m_H=\unit[137,140,170]{GeV}$ and choice II with
    $m_H=\unit[170]{GeV}$.  The middle and lower lines correspond to
    the choice II with $m_H=\unit[140]{GeV}$ and $\unit[137]{GeV}$,
    respectively.}
  \label{fig:potential}
\end{figure}

Figure \ref{fig:potential} shows the resulting RG improved potential
for several values of the Higgs mass.  It is seen, that for the choice
I the shape of the potential is nearly universal, while the overall
normalization is always the same, due to the proper choice of $\xi$.
The form of the potential (related to the $\lambda(\mu)/\xi^2(\mu)$
ratio) start to change only very close to the boundaries of the
allowed mass for the Higgs field, when the zero or the pole of
$\lambda$ are close to the inflationary value of $\mu$. For the choice
II the change of $\mu$ during the inflationary epoch is larger, so the
deviation of the potential from the tree level form is more
pronounced, especially for small Higgs masses.

\begin{figure}
  \centering
  \includegraphics[width=\columnwidth]{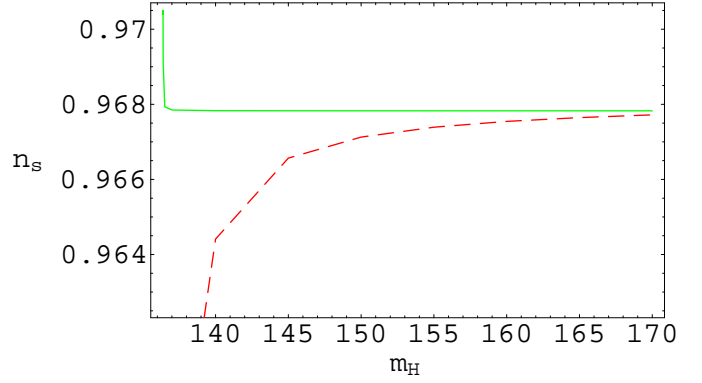}  
  \caption{Spectral index $n_s$ depending on the Higgs mass $m_H$.
    Solid line is for the choice I, dashed---for the choice II.}
  \label{fig:ns}
\end{figure}

For the choice I the spectral index stays nearly constant over the
whole admissible range of the Higgs masses, except for an extremely
small region near the boundaries (see Fig.~\ref{fig:ns}).  The
tensor-to scalar ratio $r$ also stays constantly small.  However, for
the choice II the potential is different for the lower Higgs masses,
and the spectral index becomes smaller.  At masses
$m_H<\unit[137]{GeV}$ the spectral index goes out of the WMAP allowed
region $n_s>0.93$ \cite{Komatsu:2008hk}.  Thus, the choice made in
\cite{Barvinsky:2008ia} leads, after renormalization group
improvement, to a window just slightly smaller, than the one,
following from the choice I. Moreover, contrary to what was claimed in
\cite{Barvinsky:2008ia}, the predictions of $n_s$ and $r$ depend on
the Higgs mass in the allowed interval only weakly.

Several comments are now in order. The analysis we carried out in this
Letter can be improved in several respects.\\
\begin{enumerate}
\item For the solution of the RG equations the initial conditions were
  specified with the use of the tree relations for the masses of the
  Higgs boson and t-quark. This should be modified accounting for the
  physical pole masses.
\item For the RG improvement of the potential we chose a unique scale
  $\mu$ related to the top mass (see Eq.~(\ref{renormcondition})) and
  dropped the one-loop contribution completely. This can be accounted
  for.
\item The one-loop running of the couplings can be further replaced by
  the two-loop one.
\end{enumerate}
However, these effects cannot change the main pattern of the
Higgs-inflation and will only result in some modification of the
window for the Higgs mass.

\section{Conclusions}
\label{sec:conc}

To summarize, the inflation in the Standard Model is possible for the
Higgs masses in the window $\unit[136.7]{GeV}<m_H<\unit[184.5]{GeV}$
(for $m_t = 171.2$ GeV). This roughly coincides  with the domain of
$m_H$, in which the SM can be considered as a consistent quantum field
theory all the way up to the Planck scale. For the scale invariant
normalization choice I the spectral index in the whole region is
constant and satisfies the WMAP constraints, while in the
normalization choice II from \cite{Barvinsky:2008ia} the spectral
index is $m_H$-dependent and leads to a slightly stronger limit
$m_H>\unit[137]{GeV}$ (no change in the upper limit).

If one extends the SM by three relatively light singlet fermions
($\nu$MSM of \cite{Asaka:2005an,Asaka:2005pn}), then the model
(\ref{SMJordan}) is able to address all \emph{experimentally confirmed}
indications for existence of physics beyond the SM, including neutrino
oscillations, dark matter, baryon asymmetry of the Universe, and
inflation. Further extending the model, by making it scale invariant
via introduction of one more scalar field (the dilaton)
\cite{Shaposhnikov:2008xb} and adding unimodular constraint on
gravity, allows to explain also the late time accelerating expansion
of the Universe (Dark Energy). The scale-invariant quantum
renormalization procedure of \cite{Shaposhnikov:2008xi}, applied to
this model, allows to construct a theory were all mass parameters come
from one and the same source, cosmological constant is absent due to
the symmetry requirement, and no quadratically divergent corrections to
the Higgs mass are generated. Various cosmological and experimental
consequences of the model were studied in
\cite{Asaka:2005an,Asaka:2005pn,Bezrukov:2005mx,Boyarsky:2006jm,%
Asaka:2006ek,Shaposhnikov:2006nn,Asaka:2006rw,%
Asaka:2006nq,Bezrukov:2006cy,Gorbunov:2007ak,Shaposhnikov:2007nj,%
Shaposhnikov:2008pf,Laine:2008pg,Bezrukov:2008ut,Boyarsky:2008ju,
Boyarsky:2008mx,Boyarsky:2008xj,Boyarsky:2008mt}.

These considerations indicate that no intermediate energy scale
between the $Z$ mass and the Planck scale is required to deal with the
observational and a number of fine-tuning problems of the SM\@. A
crucial test of this conjecture and of the Higgs-inflation will be
provided by LHC, if it finds nothing but the Higgs boson in a specific
mass range, found in this Letter.

A closely related paper \cite{DeSimone:2008ei} appeared in the arXiv
simultaneously with the current work, providing a different approach
to the same problem.  The conclusions of this paper about the
possibility of Higgs-inflation are similar, but not exactly identical
to ours.

The upgrade of the computation of this Letter to the two-loop level
is performed in \cite{Bezrukov:2009db}, where a more detailed
comparison with \cite{DeSimone:2008ei} is also carried out.

{\bf Acknowledgements.}

The authors thank Andrei  Barvinsky and Sergei Sibiryakov for valuable
discussions and to Andrea De Simone, Mark Hertzberg and Frank Wilczek
for exchange of information. This work was supported by the Swiss
National Science Foundation.

\bibliography{all,bookrefs}
\bibliographystyle{h-elsevier3-new}

\end{document}